\documentclass{article_saj}
\usepackage{graphicx,saj,multicol,subeqnarray}
\usepackage{natbib}
\usepackage{float}
\usepackage{xcolor}
\usepackage{widetext}
\usepackage{url}
\usepackage{bm}
\usepackage{tikz} 
\usepackage{pifont} 
\usepackage{amsfonts}
\usepackage{amssymb}
\usepackage{amsmath,upgreek}
\usepackage{titlesec}

\def\point#1{\hbox{\setbox7=\hbox to0.6em{\hfil.\hfil}%
\setbox8=\hbox to0.5em{\hfil$^{#1}$\hfil}%
\box7\kern-0.5em\box8}}

\def\pointmin#1{\hbox{\setbox2=\hbox to0.8em{\hfil.\hfil}%
\setbox3=\hbox to0.6em{\hfil$^{#1}$\hfil}%
\box2\kern-.7em\box3}}

\def\mmm{\pointmin{\mathrm{m}}\kern.15em}

\titlelabel{\thetitle.\quad}
\definecolor{xlinkcolor}{cmyk}{1,0.6,0,0}
\usepackage[bookmarks=false,         
     pdfnewwindow=true,      
     colorlinks=true,    
     linkcolor=xlinkcolor,     
     citecolor=xlinkcolor,     
     filecolor=xlinkcolor,  
     urlcolor=xlinkcolor,      
final=true
]{hyperref}

\newcommand{\OIII}{[O\,{\sc iii}]}

\usepackage[nolist]{acronym}
\begin{acronym}[AWGN]
\acro{ASKAP}{Australian Square Kilometre Array Pathfinder}
\acro{ATCA}{Australia Telescope Compact Array}
\acro{EMU}{Evolutionary Map of the Universe}
\acro{DR3}{Data Release 3}
\acro{WR}{Wolf-Rayet}
\acro{ISM}{Interstellar Medium}
\acro{WISE}{Wide-field Infrared Survey Explorer}
\acro{SNR}{Supernova Remnant}
\acro{SUMSS}{Sydney University Molonglo Sky Survey}
\acro{PMN}{Parkes-MIT-NRAO}
\acro{RSG}{Red Super Giant}
\acro{LBV}{Luminous Blue Variable}
\end{acronym}

\def\papertype{\ \hfill\ Editorial}

\def\udc{52}
\begin{document}
\parindent=.5cm
\baselineskip=3.8truemm
\columnsep=.5truecm
\newenvironment{lefteqnarray}{\arraycolsep=0pt\begin{eqnarray}}
{\end{eqnarray}\protect\aftergroup\ignorespaces}
\newenvironment{lefteqnarray*}{\arraycolsep=0pt\begin{eqnarray*}}
{\end{eqnarray*}\protect\aftergroup\ignorespaces}
\newenvironment{leftsubeqnarray}{\arraycolsep=0pt\begin{subeqnarray}}
{\end{subeqnarray}\protect\aftergroup\ignorespaces}
%


\markboth{\eightrm Evolutionary Map of the Universe: Detection of the Wolf-Rayet Star WR40} 
{\eightrm A. C. Bradley {\lowercase{\eightit{et al.}}}}

\begin{strip}

{\ }

\vskip-1cm

\publ

\type

{\ }


\title{Evolutionary Map of the Universe: Detection of the Wolf-Rayet Star WR40}


\authors{A. C. Bradley$^{1}$, Z. J. Smeaton$^{1}$, M. D. Filipovi{\' c}$^{1}$, N. F. H. Tothill$^{1}$, R. Z. E. Alsaberi$^{2,1}$}
\authors{J. D. Collier$^{3,4,1}$, Y. A. Gordon$^{5}$, A. M. Hopkins$^{6}$, and H. Zakir$^1$}


\vskip3mm


\address{$^1$Western Sydney University, Locked Bag 1797, Penrith South DC, NSW 2751, Australia}
\address{$^2$Faculty of Engineering, Gifu University, 1-1 Yanagido, Gifu 501-1193, Japan}
\address{$^3$Australian SKA Regional Centre, Curtin Institute of Radio Astronomy (CIRA), 1 Turner Avenue, Technology Park, Bentley, Western Australia, 6102}
\address{$^4$The Inter-University Institute for Data Intensive Astronomy (IDIA), Department of Astronomy, University of Cape Town, Private Bag X3, Rondebosch, 7701, South Africa}
\address{$^5$Department of Physics, University of Wisconsin-Madison, 1150 University Avenue, Madison, WI 53706, USA}
\address{$^6$School of Mathematical and Physical Sciences, 12 Wally’s Walk, Macquarie University, NSW 2109, Australia}

\Email{20295208@student.westernsydney.edu.au}


\dates{XXX XX, 2025}{XXX XX, 2025}


\summary{We present a radio-continuum detection of the well-known Wolf-Rayet star WR40 at 943.5\,MHz using observations from the \ac{EMU} survey. We find that the shell surrounding WR40, known as RCW~58,  has a flux density of 158.9$\pm$15.8\,mJy and the star itself is 0.41$\pm$0.04\,mJy. The shell size is found to be 9$^{\prime}$$\times$~6$^{\prime}$, which matches well with the shell in H$\alpha$ and is similarly matched to the shell at 22\,$\mu$m in infrared. Using \textit{Gaia} data, we derive a linear size of $7.32(\pm0.34)\times4.89(\pm0.23)$\,pc at a distance of 2.79$\pm$0.13\,kpc. We use previous \ac{ATCA} observations at 8.64, 4.80, and 2.4\,GHz to determine a spectral index of WR40, which is estimated to be $\alpha\,=\,0.80\pm0.11$, indicating that the emission from the star is thermal.}


\keywords{Radio continuum: stars --- Stars: Wolf-Rayet
 --- ISM: individual objects: WR40 --- Astrometry}

\end{strip}

\tenrm


\section{Introduction}
\label{sec:intro}
\indent

\ac{WR} stars are late-stage stars, and are often defined by their mass loss rate compared to previous evolutionary stages \citep{2000A&A...360..227N}, creating outbursts of stellar material mainly comprised of the star's dominating element \citep{1995AJ....109.1839M}. They are characterised by the elements dominating their spectra; either carbon, nitrogen, or oxygen \citep{2007ARA&A..45..177C}. The elemental composition of \ac{WR} stars categorises them into three main types: WC (carbon-rich), WN (nitrogen-rich), and WO (oxygen-rich). 

WR40 is a well-known WN8 type \ac{WR} star \citep{2006A&A...457.1015H} accompanied by a shell known as RCW~58, which has also been studied extensively \citep{1986MNRAS.221..715H}. RCW~58 is unique, possessing a non-uniform, irregular nature, likely caused by interactions of the wind-driven ejecta with the \ac{ISM} \citep{1986MNRAS.221..715H, 1995AJ....109.1839M, 2021MNRAS.507.3030J}. WR40 is often linked to  WR16 \citep{1995AJ....109..817A, 2020MNRAS.495..417C,2025PASA...42..101B}, another WN8 star. They are similar, but WR40's shell is quite irregular and elliptical, whereas WR16's is symmetrical and circular \citep{1995AJ....109.1839M}. WR stars are known to have variability, with WR40 showing significant variability compared to others \citep{1989MNRAS.238...97G}.



The \ac{EMU} survey \citep{2025PASA...42...71H,Norris2011,Norris2021} is mapping the entire southern sky at 943.5\,MHz using the \ac{ASKAP} \citep{2021PASA...38....9H}. Due to the telescope's high sensitivity, we often find low-surface brightness features never before characterised. Some of these objects are \acp{SNR};
G305.4-2.2~\citep[Teleios;][]{2025PASA...42..104F}, J0624--6948~\citep{2022MNRAS.512..265F,2025A&A...693L..15S}, G288.8--6.3~\citep[Ancora;][]{2023AJ....166..149F,2024A&A...684A.150B}, G308.7+1.4 \citep[Raspberry;][]{2024RNAAS...8..107L}, G312.6+ 2.8 \citep[Unicycle;][]{2024RNAAS...8..158S}; a pulsar wind nebula (PWN) \citep[Potoroo;][]{2024PASA...41...32L}; a reflection nebula (RNe) \citep{2025PASA...42...32B}; and possible candidates for the source of the Ultra-high-energy Neutrino Event KM3{\textendash}230213A \citep{2025ApJ...984L..52F}.


\section{Data}
\label{sec:data}
\indent

\subsection{ASKAP EMU}
\label{Sub_Section:Data:ASKAP_EMU}
\indent

\begin{figure}[!ht]
\centering
    \includegraphics[width=\linewidth]{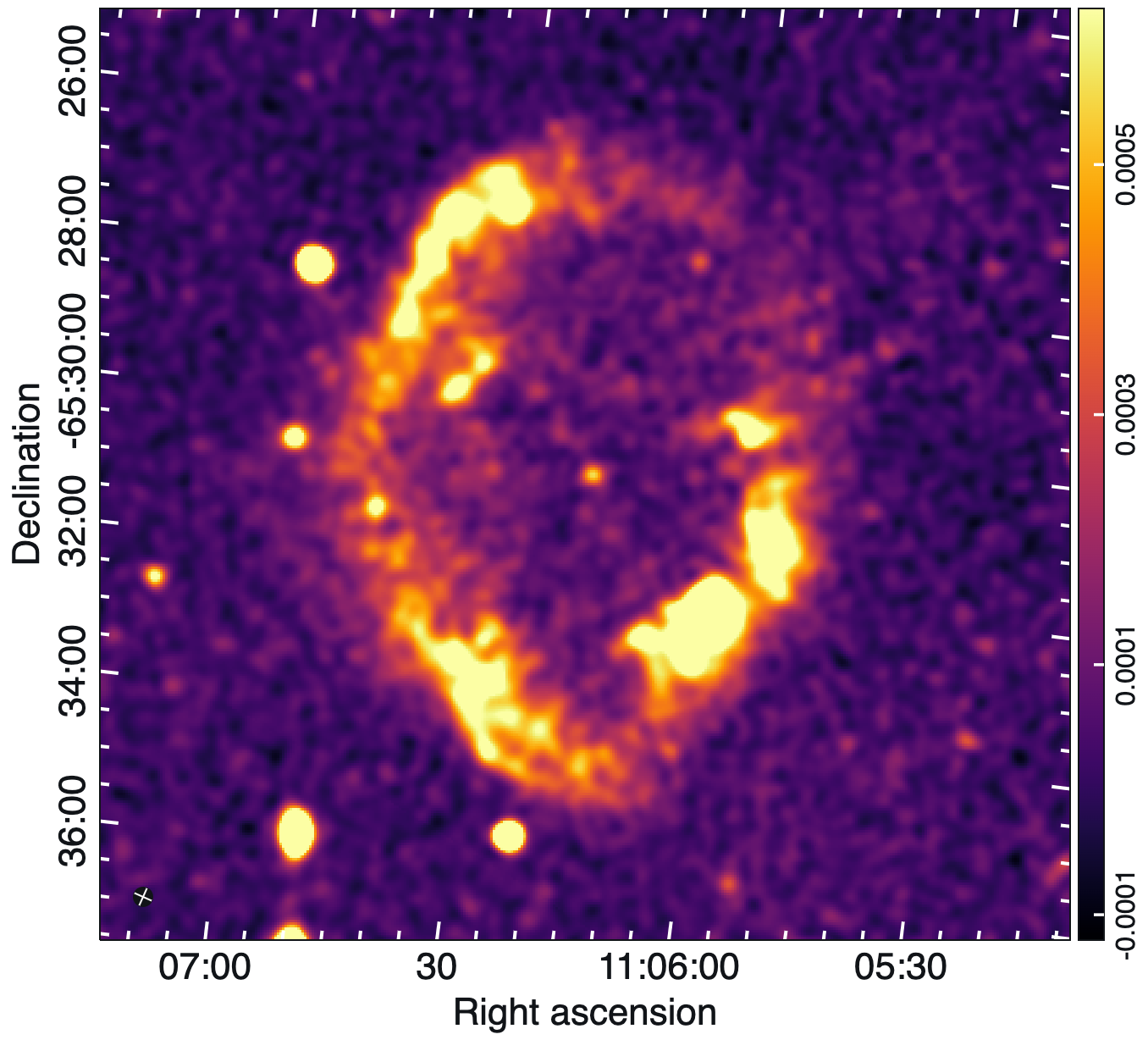}
    \caption{\ac{EMU} 943.5\,MHz radio-continuum image of WR40 and its surrounding shell RCW~58 at 15$^{\prime\prime}$ resolution. Image is linearly scaled.}
    \label{fig:Radio-Standalone}
\end{figure}

WR40 and its shell have been seen in two \ac{EMU} observations. SB46948 observed the tile EMU$\_1136$-$64$ on December 12 2022, and SB54771 observed the tile EMU$\_1050$-$64$ on November 11 2023. These observations were reduced using the standard \ac{ASKAP} pipeline, ASKAPSoft, using multi-frequency synthesis imaging, multi-scale cleaning, self-calibration and convolution to a common beam size \citep{2019ascl.soft12003G}.
We then combined the observations into a single image using the Miriad~\citep{miriad} task \textsc{imcomb}. The final image was created using equal weighting, resulting in a lower root mean square (RMS) noise level of $\sim$30$\,\mu$Jy beam$^{-1}$. The final image is shown in Figure~\ref{fig:Radio-Standalone}. 


\subsection{Other Data}
\label{Sub_Section:Data:Other_Data}
\indent

We use values from the \textit{Gaia} \ac{DR3} catalogue observations \citep{2016A&A...595A...1G,2023A&A...674A...1G} to determine the distance to WR40, as well as determining the true size of the radio shell surrounding it. 

\begin{figure*}[!ht]
\centering
    \includegraphics[width=\linewidth]{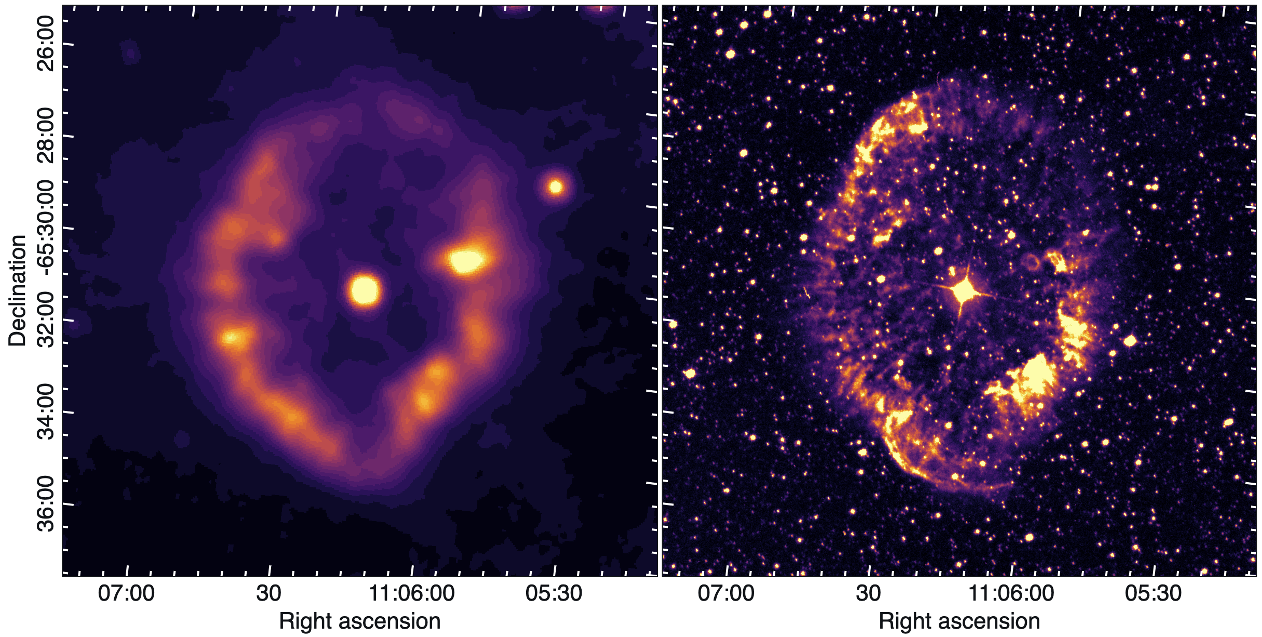}
    \caption{WR40 and its surrounding shell RCW~58, both images are linearly scaled. --  \textbf{Left:} WISE infrared image in the W4 band at 22\,$\mu$m. -- \textbf{Right:} SuperCOSMOS H$\alpha$ (656.281~nm) image.}
    \label{fig:WR16/WR40}
\end{figure*}

We also include observations from the SuperCOSMOS \citep[656.281~nm][]{2001MNRAS.326.1279H} and \ac{WISE} \citep[22~$\mu$m][]{2010AJ....140.1868W} sky surveys to compare the nebulosity surrounding WR40 at different wavelengths (Figure~\ref{fig:WR16/WR40}). 

The central star WR40 is seen in RACS-High \citep[1655.5~MHz][]{2021PASA...38...58H} as measured in the Sydney Radio Stars Catalogue \citep{2024PASA...41...84D}. It is also measured at 943.5~MHz, but because the integrated flux is much higher than our \ac{EMU} flux, we have elected to use neither of the catalogue's measurements in our analysis.


\section{Results and Discussion}
\label{sec:results}
\indent

In relation to WR40, there are disparities between parallax values in previous literature \citep{1997A&A...323L..49P, 2007A&A...474..653V, 2016A&A...595A...2G}. We use the \textit{Gaia} Data Release 3 (DR3) parallax of 0.3572$\pm$0.0170\,milliarcseconds \citep{2020yCat.1350....0G} to estimate a distance for WR40 to be 2.79$\pm$0.13\,kpc. We use this distance and the angular diameter measured from the \ac{EMU} image (9$^{\prime}$$\times$~6$^{\prime}$) to determine the true size of the shell to be $7.32\pm0.34\times4.89\pm0.23$\,pc. Within this defined region, we measure the flux density of the WR40 nebulosity to be 158.9$\pm$15.8\,mJy, taking a 10\% error \citep{2022MNRAS.512..265F,2024PASA...41..112F}. We also measure WR40 itself and find its flux density to be 0.41$\pm$0.04\,mJy. 

Using previous \ac{ATCA} measurements, we can determine a spectral index for the star WR40. \citet{1995ApJ...450..289L} determines a flux density of 2.52$\pm$0.09\,mJy at 8.64\,GHz, and 1.69$\pm$0.10\,mJy at 4.80\,GHz. \citet{1999ApJ...518..890C} measured the star WR40 at 2.40\,GHz, estimating a flux density of 1.21$\pm$0.13\,mJy. The spectral index is defined as $S \propto \nu^{\alpha}$, where $S$ is flux density, $\nu$ is the frequency and $\alpha$ is the spectral index \citep{book1}. Using the three ATCA measurements and our \ac{EMU} measurement, we estimate a spectral index of $\alpha\,=\,0.80\pm0.11$, indicating thermal emission from the star. We obtain a consistent measurement using only the 1999 observations, suggesting radio variability isn't significantly impacting our spectral index measurement, despite WR40 showing variability at other wavelengths \citep{1989MNRAS.238...97G}.

\begin{figure*}[h]
\centering
    \includegraphics[width=\linewidth]{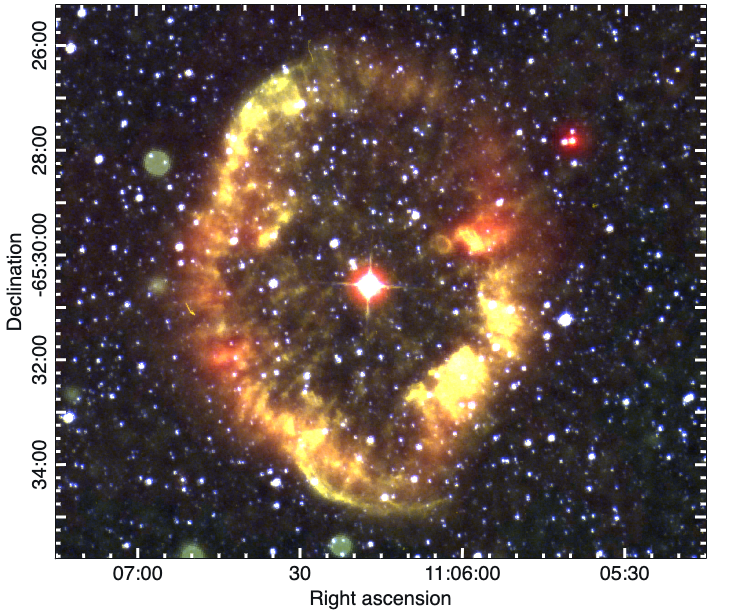}
    \caption{Four-colour composite image of WR40 and RCW~58. Red is \ac{WISE} 22~$\mu$m, yellow is SuperCOSMOS H$\alpha$, green is the \ac{EMU} observation, and blue is the DSS2 red plate provided from \citet{1996ASPC..101...88L}.}
    \label{fig:RGB}
\end{figure*}

The emission seen at 943.5\,MHz matches well with SuperCOSMOS H$\alpha$ and \ac{WISE} 22\,$\mu$m (Figure~\ref{fig:RGB}), which may indicate that the radio emission is thermal. We have attempted to obtain additional radio data in order to constrain the nature of the emission of the shell, measuring fluxes from the \ac{SUMSS} \citep{1999AJ....117.1578B} and \ac{PMN} survey \citep{1993AJ....105.1666G}. We also obtained flux measurements detailed in \citet{2021MNRAS.507.3030J} from \ac{ATCA} observations. However, we were unable to determine a reliable spectral index with any of these observations due to the changes in sensitivity and the low surface brightness of the source.

\citet{2019ApJ...884L..49P} explored the non-thermal nature of a WO-type \ac{WR} star with a spectral index of $-0.81\pm 0.1$. We have applied this spectral index to our \ac{EMU} measurement to predict a \ac{SUMSS} flux at 843~MHz. We predict an expected flux of $\sim$173~mJy, as this is brighter than the \ac{EMU} emission, we would expect a reliable detection. However, WR40 as seen by \ac{SUMSS} is almost indistinguishable from the background noise. This makes the thermal scenario more likely, however there is not enough evidence to rule out the non-thermal scenario. It is possible that follow up observations with a sensitive telescope like MeerKAT \citep{2016mks..confE...1J} could help constrain the nature of the emission.


\citet{1995AJ....109.1839M} shows observations of \OIII\ which extends outside the radio shell toward the south-east portion which does not appear in the \ac{EMU} observation, but is partially seen in the \ac{WISE} image. Differences in the shell of the three observations may be partially explained by differences in resolution, however there are some noticeable bright spots present in the \ac{WISE} (Figure~\ref{fig:WR16/WR40}, Left) observation that are not present in the \ac{EMU} and SuperCOSMOS H$\alpha$ observations. Due to the irregular shape of RCW~58, it is likely that the wind from WR40 is expanding into a slower, non-uniform wind from the previous \ac{RSG} or \ac{LBV} phase (\citep{2021MNRAS.507.4697M})



\section{Conclusion}
\label{sec:conclusion}
\indent

We present a detection of the well known Wolf-Rayet star WR40 and its shell RCW~58 at 943.5~MHz using observations from the \ac{EMU} survey. The shell as seen by \ac{EMU} is 9$^{\prime}$$\times$~6$^{\prime}$, and using \textit{Gaia} data, we derive a true size of $7.32(\pm0.34)\times4.89(\pm0.23)$\,pc at a distance of 2.79$\pm$0.13\,kpc. We measure a flux density for RCW~58 to be 158.9$\pm$15.8\,mJy, and WR40's flux density as 0.41$\pm$0.04\,mJy. Using previous \ac{ATCA} observations, at 8.64, 4.80, and 2.4\,GHz, we calculate a spectral index for the star WR40 of $\alpha\,=\,0.80\pm0.11$, indicating that the emission from the star is thermal. We are unable to determine a spectral index for RCW~58 due to its low surface brightness and inconsistencies in previous radio observations. Taking observations with telescopes from the latest generation such as MeerKAT will help constrain the spectral index and determine the nature of the emission of RCW~58.



\acknowledgements{We thank Bärbel S. Koribalski for helpful discussions. 
This scientific work uses data obtained from Inyarrimanha Ilgari Bundara, the CSIRO Murchison Radio-astronomy Observatory. We acknowledge the Wajarri Yamaji People as the Traditional Owners and native title holders of the Observatory site. CSIRO’s ASKAP radio telescope is part of the Australia Telescope National Facility (\url{https://ror.org/05qajvd42}). Operation of ASKAP is funded by the Australian Government with support from the National Collaborative Research Infrastructure Strategy. ASKAP uses the resources of the Pawsey Supercomputing Research Centre. Establishment of ASKAP, Inyarrimanha Ilgari Bundara, the CSIRO Murchison Radio-astronomy Observatory and the Pawsey Supercomputing Research Centre are initiatives of the Australian Government, with support from the Government of Western Australia and the Science and Industry Endowment Fund.
This work has made use of data from the European Space Agency (ESA) mission {\it Gaia} (\url{https://www.cosmos.esa.int/gaia}), processed by the {\it Gaia} Data Processing and Analysis Consortium (DPAC, \url{https://www.cosmos.esa.int/web/gaia/dpac/consortium}). Funding for the DPAC has been provided by national institutions, in particular the institutions participating in the {\it Gaia} Multilateral Agreement.}



\newcommand\eprint{in press }

\bibsep=0pt

\bibliographystyle{aa_url_saj}

{\small

\bibliography{WR40}
}

\begin{strip}

\end{strip}

\clearpage

{\ }


\end{document}